\documentclass[pre,aps,twocolumn,superscriptaddress,amsmath,showpacs,floatfix]{revtex4}

\usepackage{amsmath}
\usepackage{setspace}
\usepackage{epsfig}
\usepackage{color}

\usepackage{amssymb}
\begin{document}
\title{Small membranes under negative surface tension} \author{Yotam
Y. Avital and Oded Farago\\ Department of Biomedical Engineering and
Ilse Katz Institute for Nanoscale\\ Science and Technology, Ben-Gurion
University of the Negev,\\ Be'er Sheva 84105, Israel.}

\begin{abstract} 

We use computer simulations and a simple free energy model to study
the response of a bilayer membrane to the application of a negative
(compressive) mechanical tension. Such a tension destabilizes the long
wavelength undulation modes of giant vesicles, but it can be sustained
when small membranes and vesicles are considered. Our negative tension
simulation results reveal two regimes~-~(i) a weak negative tension
regime characterized by stretching-dominated elasticity, and (ii) a
strong negative tension regime featuring bending-dominated elastic
behavior.  This resembles the findings of the classic Evans and Rawicz
micropipette aspiration experiment in giant unilamellar vesicles
(GUVs) {[}Phys, Rev. Lett.  {\bf 64}, 2094 (1990){]}.  However, while
in GUVs the crossover between the two elasticity regimes occurs at a
small positive surface tension, in smaller membranes it takes place at
a moderate negative tension. Another interesting observation
concerning the response of a small membrane to negative surface
tension is related to the relationship between the mechanical and
fluctuation tensions, which are equal to each other for non-negative
values.  When the tension decreases to negative values, the
fluctuation tension $\gamma$ drops somewhat faster than $\tau$ in the
small negative tension regime, before it saturates (and becomes larger
than $\tau$) for large negative tensions. The bending modulus exhibit
an ``opposite'' trend. It remains almost unchanged in the
stretching-dominated elastic regime, and decreases in the
bending-dominated regime. Both the amplitudes of the thermal height
undulations and the projected area variations diverge at the onset of
mechanical instability.

\end{abstract}
\maketitle 

\vspace{0.45cm}

\newpage

\section{Introduction}
\label{sec:intro}

Bilayer membranes are quasi-two-dimensional (2D) fluid sheets formed
by spontaneous self-assembly of lipid molecules in water
\cite{israelachvili}. Their elasticity is traditionally studied in the
framework of the Helfrich effective surface Hamiltonian for 2D
manifolds with local principle curvatures $c_1$ and $c_2$
\cite{helfrich:73}
\begin{equation}
{\cal H}=\int_{A} dS
\left[\sigma_0+\frac{1}{2}\kappa_0
\left(c_1+c_2-2c_{0}\right)^2 
+{\bar{\kappa}}_0c_1c_2\right]
\label{eq:helfhamiltonian}
\end{equation}
where the integration is carried over the whole surface of the
membrane. The Helfrich Hamiltonian involves four parameters: the
spontaneous curvature $c_0$, the surface tension $\sigma_0$, the
bending modulus $\kappa_0$, and the saddle-splay modulus
$\bar{\kappa}_0$. For symmetric bilayer membranes, $c_0=0$. If, in
addition, the discussion is limited to deformations that preserve the
topology of the membrane, then (by virtue of the Gauss-Bonnet theorem)
the total energy associated with the last term is constant, and one
arrives to the more simple form
\begin{equation}
{\cal H}=\int_{A} dS
\left[\sigma_0+\frac{1}{2}\kappa_0\left(c_1+c_2\right)^2\right]
=\sigma_0 A+\frac{1}{2}\kappa_0 J^2,
\label{eq:helfhamiltonian2}
\end{equation}
where $A$ is the total area of the membrane and $J$, defined by
$J^2=\int dS \left(c_1+c_2\right)^2$, is the integrated total
curvature.
 
Helfrich Hamiltonian provides successful framework for describing many
features of bilayer membranes and vesicles, including their
large-scale shapes and transformations between them
\cite{seifert_rev}, membrane-membrane interactions
\cite{lipowsky_rev}, and membrane-mediated forces between proteins
(``inclusions'') \cite{bruinsma_rev}.  In light of this success and
the wide acceptance off the model within the soft matter and
biophysics communities, it is surprising how poorly understood and
confusing remains the surface tension term $\sigma_0$ in
Eq.~(\ref{eq:helfhamiltonian2}). Below we briefly review some of the
complications associated with the concept of membrane surface tension.

1. Membranes are much more flexible to bending than stretching.
Therefore, in theoretical studies it is sometimes assumed that the
lipids area density is fixed \cite{sens}, and changes in the total
area result in from changes in the number of lipids \cite{david}. In
this picture, the surface tension is essentially the chemical
potential of the surface lipids \cite{farago:2003}.  Lipids, however,
are highly insoluble in water (their critical micelle concentration is
typically in the $<10^{-6}$ M range \cite{israelachvili}), which makes
the above interpretation for the surface tension largely irrelevant as
there is almost no exchange of lipids between the bilayer membrane and
the embedding aqueous medium. More commonly, the surface tension is
considered as a measure for the elastic response of a membrane with a
fixed number of lipids to area variations \cite{schulman,
  taupin}. This view clearly differs from the meaning of the term
``surface tension'' for fluid/fluid interfaces, where it serves as a
measure for the free energy penalty resulting from an exchange of
molecules between the ``bulk'' phases and the ``interface''
\cite{rowlinson}.

2. Further complication arises from the fact that the membrane is a
(quasi) two dimension manifold embedded in a three dimensional
space. Therefore, it has two characteristic areas - (i) the total
physical area $A$, and (ii) the area of its projection onto a planar
reference frame, $A_p$. Helfrich Hamiltonian involves an integral over
the total area $A$; but this area (whose determination, taking into
account the molecular structure of the membrane, is not without
ambiguity) cannot be fixed due to the membrane thermal
undulations. Thus, $A$ is not a valid thermodynamic variable
\cite{diamant}. It is, in fact, the projected area $A_p$ that emerges
as the computationally relevant quantity \cite{comment4}. The force
per unit length that needs to imposed on the frame in order to fix its
area to $A_p$ is known as the {\em mechanical (frame)}\/ tension, and
will be denoted henceforth by $\tau$.  Theoretically, the mechanical
tension can be identified with the derivative of the elastic free
energy, $F$, with respect to $A_p$: $\tau=\partial F/\partial A_p$
(where the differentiation is carried while holding the volume of the
system, as well as the temperature and number of lipids, fixed).  One
can also fix $\tau$ and let $A_p$ fluctuate. In this case, the
relevant free energy is $G=F-\tau A_p$, and the mean projected area
satisfies $\langle A_p\rangle=-\partial G/\partial \tau$.

3. Another confusing aspect associated with the concept of surface
tension is the distinction between Helfrich Hamiltonian and the
membrane free energy $F$.  The latter is often referred to as the {\em
  Helfrich free energy}\/, and the confusion arises because it is
assumed to have the same form as the Helfrich Hamiltonian
\begin{eqnarray}
F=\sigma \bar{A}+\frac{1}{2}\kappa\bar{J}^2.
\label{eq:helffreee}
\end{eqnarray}
In Eq.~(\ref{eq:helffreee}), $\bar{A}$ and $\bar{J}$ denote,
respectively, the area and total curvature of the {\em mean profile}\/
of the membrane. Eq.~(\ref{eq:helffreee}) features two {\em
renormalized}\/ coefficients, the tension $\sigma$ and bending modulus
$\kappa$, that are not equal to their {\em intrinsic}\/ counterparts
from Eq.~(\ref{eq:helfhamiltonian2}), $\sigma_0$ and $\kappa_0$.  The
statistical mechanics of thermal fluctuations around the mean profile
are accounted for in the values of the renormalized elastic
coefficients.  For a membrane with a mean flat profile (i.e., not
subjected to normal bending forces) $\bar{A}=A_p$ and,
$\bar{J}=0$. Then, Eq.~(\ref{eq:helffreee}) takes the form $F=\sigma
A_p$, and $\tau=\partial F/\partial A_p=\sigma$.

4. Another quantity that can be identified as the membrane surface
tension is the, so called, {\em $q^2$}-coefficient\/ $\gamma$, also
known as the {\em fluctuation}\/ tension. The fluctuation tension is
measured from the Fourier spectrum of the membrane height function
with respect to the plane of the frame.  For a membrane with a mean
flat profile, the thermal average of the amplitude of a Fourier mode
with wavevector $\vec{q}$ satisfies: $\langle h_{\vec{q}}\rangle=0$,
and
\begin{equation}
\left\langle |h_{\vec{q}}|^2 \right\rangle=\frac{k_BT A_p}{l^4\left[\gamma
q^2 +\kappa q^4 +{\cal O}(q^6)\right]},
\label{eq:flucspect1}
\end{equation}
where $k_B$ is Boltzmann constant, $T$ is the temperature, and $l$ is
a microscopic cutoff length. Some controversy surrounded
Eq.~(\ref{eq:flucspect1}) concerning the question whether $\gamma$,
the $q^2$-coefficient in the denominator on the right hand side, is
equal to $\sigma_0$ or $\sigma$. This issue has been recently settled,
and it now understood that the correct coefficient is the renormalized
surface tension $\sigma$ \cite{farago04,farago11}, or, according to some
theoretical studies, a slightly modified version of this surface
tension, $(A_p/A)\sigma$ \cite{schmid}. Similarly, the $q^4$
coefficient in Eq.~(\ref{eq:flucspect1}) is equal to the renormalized
bending modulus $\kappa$. Combining this with the discussion about
Eq.~(\ref{eq:helffreee}), we conclude that both the measurable
fluctuation and mechanical tensions coincide with each other (and with
the renormalized tension $\sigma$): $\gamma=\tau=\sigma$.

\section{Negative surface tension}
\label{sec:negten}

When the surface tension $\tau=\gamma=\sigma$ vanishes, the membrane
is ``free to choose'' the equilibrium projected area $A_p$ that
minimizes the free energy $G=F$.  The question we now wish to address
concerns with the elastic response of the membrane to a further
decrease in the frame area, which involves the application of a
negative surface tension. Based on Eq.~(\ref{eq:flucspect1}), one may
argue that for $\gamma<0$, the membrane always becomes mechanically
unstable because the amplitude of any mode with
$q<\sqrt{(-\gamma/\kappa)}$ diverges. But such modes exist only in
sufficiently large membranes, hence, small membranes can always
sustain some negative surface tension.  For instance, consider a
square membrane of linear size $L$ with $\kappa=25k_BT\simeq
10^{-19}{\rm J}$.  From Eq.~(\ref{eq:flucspect1}), one finds that such
a membrane can withstand negative surface tension of size
$\gamma=-5\times10^{-3}{\rm N/m}$ (which is comparable in magnitude to
the typical positive rupture tension), provided that
$L<(2\pi)\sqrt{-\kappa/\gamma}\sim 30{\rm nm}$. This is the
characteristic size of real small liposomes, and of bilayers in highly
coarse-grained simulations.  Thus, the above estimation highlights the
fact that the question of elastic response to negative surface tension
is both interesting and relevant to current experimental and
computational studies.

The derivation of Eq.~(\ref{eq:flucspect1}), and the proof that the
fluctuation and mechanical tensions coincide with each other, involves
several assumptions that do not necessarily remain valid when $\sigma$
becomes negative.  Specifically, it is based on the investigation of
the linear response of a mechanically {\em stable}\/ flat membrane to
small normal forces, and is restricted to configurations with smooth
(twice differentiable) height functions $h(\vec{r})$. But when
$\sigma<0$, the membrane can relieve the free energy cost of
compression by buckling. We note that {\em in the absence of normal
forces}\/ (which is the case under consideration here), the system is
not expected to undergo spontaneous symmetry breaking similar to that
occurring e.g., in Ising spin systems below the critical point. The
reason is that the membrane height profile $h(\vec{r})$ is a
continuous field and, therefore, the transition between different
buckled configurations (e.g., from buckled ``upward'' to ``downward'')
does not require the crossing of a free energy barrier. Thus, the
system remains ergodic for negative tension, and due to the symmetry
of the bilayer $\langle h_{\vec{q}}\rangle=0$ for all the Fourier
modes. The questions that remain are:

1. Does Eq.~(\ref{eq:flucspect1}) still hold for $\sigma<0$? We
certainly do not expect it to remain valid for strongly compressed
membranes since it is derived from a quadratic (in the height
function) approximation to the full Helfrich Hamiltonian
(\ref{eq:helfhamiltonian2}). However, considering the fact that it
holds for $\sigma=0$, there is no apparent reason why it should not
hold for small negative $\sigma$.

2. Are the mechanical and fluctuation tensions still equal to each
other? (Obviously, this question is relevant only if the answer to
question no.~1 is ``yes''.)  As noted above, the proof of this
equality depends on the surface tension being positive. Now that it is
negative, the membrane prefers more buckled configurations with larger
mean squared amplitudes. Does this imply that the fluctuation tension
$\gamma$ drops faster (i.e., becomes more negative) than the
mechanical tension $\tau$?

3. What happens to the bending modulus $\kappa$ under compression?
Recall that the coefficient appearing in Eq.~(\ref{eq:flucspect1}) is
the renormalized bending modulus which, just like the tension
$\gamma$, may vary with the frame area $A_p$. For positive tensions,
the variations in $\kappa$ are usually negligible, but this may not be
the case for negative tensions when the membrane becomes increasingly
more buckled. Does the increase in the degree of buckling under larger
compressive stresses involve a decrease in $\kappa$?

4. Does the membrane exhibit linear (Hookean) elastic response to
   negative mechanical tension?  In response to a positive tension,
   the membrane becomes stretched, and the relationship between the
   change in the area (strain) and the stress is indeed
   linear. However, the lipids constitute a dense two-dimensional
   fluid and, therefore, the membrane can be barely compressed below
   its most favorable physical area $A_0$. When, under the application
   of a negative tension, the physical area $A$ reaches $A_0$, the
   negative tension causes the membrane to buckle and more and more
   area is ``stored'' in the out-of-plane fluctuations. This could
   lead to a highly non-linear elastic response.

In what follows we will use Monte Carlo (MC) simulations to address
the above questions.

\section {Computer simulation methods}
\label{sec:methods}

To allow for large scale membrane simulations ($N=1000$ lipids per
monolayer) in a computationally feasible manner, we employ the
Cooke-Deserno model \cite{deserno:2005}, with model parameters used in
our recent work on fluid charged membranes \cite{Farago:2014}. In this
highly coarse-grained model, the lipids are represented as trimers
with one hydrophilic and two hydrophobic beads, and the embedding
solvent is handled implicitly via effective interactions between the
hydrophobic beads. The membranes are simulated in a box of linear size
$L_x=L_y=L$, with periodic boundary conditions in the $x-y$ (frame)
plane. Initially, we place a pre-assembled flat membrane at the center
of the simulation box, and we then allow it to equilibrate for $1
\times 10^5$ MC time units. On average, each MC time unit consists of
$N$ translation (with additional small intramolecular displacements)
and $N$ rotation move attempts carried on randomly chosen lipids. The
membrane is simulated at constant frame tension $\tau$, which is
accomplished by incorporating several move attempts, per time unit, to
change the frame area of the membrane \cite {Farago:2007}.  Each time
unit also includes several collective ``mode excitation'' moves
\cite{Farago:2008} that accelerate the slow dynamics of the
long-wavelength Fourier modes. For each value of $\tau$, the initial
thermalization period is followed by a period of $1.5 \times 10^6$ MC
time units during which we sample quantities of interest at 200 MC
units intervals.

In what follows, we set the thermal energy $k_BT$ to be the elementary
energy unit, and the length parameter of the Deserno-Cooke repulsive
potential $b$ to be the length unit. We simulate the membrane under
frame tensions satisfying $-0.3\leq\tau\leq 0.5$ (in $k_BT/b^2$ units)
which, as revealed by our computational observations, is the stability
range of the simulated membranes. Relation to physical units can be
made by setting $b=0.65$ nm, which gives the unit of the surface
tension $k_BT/b^2\simeq 10$ mN/n. For $\tau>0.5$, the membranes
rupture, while for $\tau<-0.3$, they exhibit large normal undulations
leading to collapse and dissociation of lipids. Within the stability
range, we measure the mean and variance of the projected area
distribution ($\left \langle A_p \right \rangle$ and $\left \langle
\delta A_p^2 \right \rangle = \left \langle A_p^2 \right \rangle -
\left \langle A_p \right \rangle ^2$, respectively). We also calculate
the Fourier transform of the height undulations, by dividing the
membrane into $8\times 8$ grid cells, and calculating the local mean
height of the bilayer within each grid cell.  The Fourier transform of
$h(\vec{r})$ in wavenumber space $\vec{n}=\vec{q}(L/2\pi)$
($\vec{n}=(n_x,n_y)\ ; \ n_x,n_y=-4,-3,\ldots,2,3$) is defined by

\begin{eqnarray}
\tilde h_{\vec{n}}=\frac{1}{L} \sum_{\vec{r}} h(\vec{r})e^{-2\pi i
\vec{n}\cdot\vec{r}/L}.
\label{eq:fourier}
\end{eqnarray}
Notice that the linear size of the frame $L$, appearing (twice) in the
definition of $h_{\vec{n}}$, is not constant, but rather fluctuates
during the course of the (constant tension) simulations. At each
measurement, we use the instantaneous value of $L$. Also notice that
$h_{\vec{n}}$ defined in Eq.~(\ref{eq:fourier}) is dimensionless, due
to the $L^{-1}$ prefactor that does not exist in the more commonly
used $h_{\vec{q}}$ of Eq.~(\ref{eq:flucspect1}).  In terms of the
variable $h_{\vec n}$, Eq.~(\ref{eq:flucspect1}) takes the form
\begin{equation}
\left\langle |h_{\vec{n}}|^2
\right\rangle=\left(\frac{L}{l}\right)^4\frac{k_BT}{\left[\gamma
\langle A_P\rangle (2\pi n)^2 +\kappa (2\pi n)^4 \right]}.
\label{eq:flucspect2}
\end{equation}

There are four different modes corresponding to each value of
$|\vec{n}|$, and this number is reduced to two if $|n_x| = |n_y|$, or
if one of the components of $\vec{n}$ is zero. In the following
section, the results for $ \left\langle |h_{\vec{n}}|^2 \right\rangle$
(and other related quantities) represent averages over these distinct
modes. In Eq.~(\ref{eq:flucspect2}), $l$ is the grid size, which
implies that $L/l=8$, independently of the instantaneous value of~$L$.

\begin{figure}[!t]
  \centering
  \includegraphics[width=0.5\textwidth]{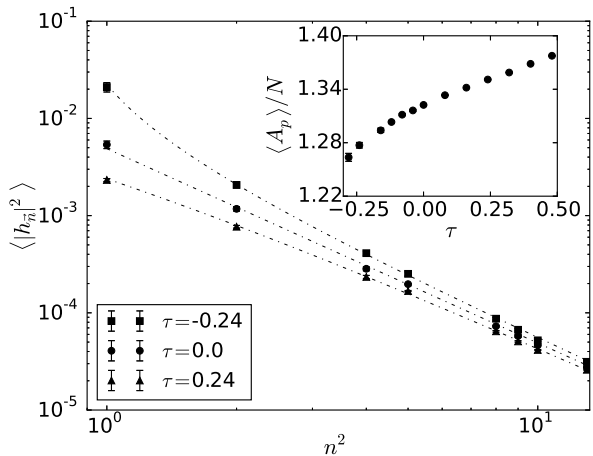}
  \caption{The spectral intensity as a function of the wavenumber for
    membranes under frame tension of $\tau=-0.24$ (squares), 0
    (circles), and 0.24 (triangles). The dotted-dashed curves represent
    the best fits of the results to Eq.~(\ref{eq:flucspect2}) over the
    first four modes. The inset shows the mean projected area per
    lipid as a function of $\tau$.}
  \label{fig:flucspect2}
  \vspace{-0.18cm}
\end{figure}

Due to molecular-scale protrusion, the physical area of the membrane
cannot be unambiguously determined. Therefore, we use the following
approximation for $\langle A\rangle$
\begin{equation}
  \left \langle A \right \rangle \simeq  \left \langle A_p \right \rangle
  \left [ 1+ \frac{1}{2}\left(\frac{l}{L}\right)^4
\sum_{\vec n} (2\pi n)^2 \left \langle \left |
  \tilde h_{\vec{n}} \right|^2 \right \rangle\right],
\label{eq:totalarea}
\end{equation}
which is the physical area ``visible'' up to the resolution of the
grid.  One can also define the effective area-stretch modulus of the
membrane, $K_A$, by assuming that the free energy cost due to small
variations in the projected area from $\langle A_p\rangle$ can be
approximated by the quadratic form $(1/2)K_A[A_p-\langle
A_p\rangle]^2/\langle A_p\rangle$ \cite{farago:2003}. Under this
approximation, the coefficient $K_A$ can be extracted from the
fluctuation statistics of $A_p$ using the equipartition theorem
\begin{equation}
  K_A = \frac{k_BT\left\langle A_p \right\rangle }{ \left \langle
  \delta A_p^2 \right \rangle}.
  \label{eq:compress1}
\end{equation}

\section{Results}
\label{sec:results} 

In section \ref{sec:negten} we raised several questions concerning the
elastic and fluctuation behavior of membranes under negative
mechanical tension. Here, we present computer simulations results
addressing those question.

We begin with the question of the validity of
Eq.~(\ref{eq:flucspect2}) for negative frame tensions. Figure
\ref{fig:flucspect2} displays our results for the fluctuation spectral
intensity, $|h_{\vec{n}}|^2$, as a function of $n^2$ for membranes
under three different mechanical tensions $\tau=-0.24,\ 0,\ 0.24$.
The fits of the computational results to Eq.~(\ref{eq:flucspect2}) are
displayed with dotted-dashed lines. Our analysis reveals that within
most of the range of mechanical stability $-0.3 \leq \tau \leq 0.5$,
the quality of each fit is very good. This demonstrates that
Eq.~(\ref{eq:flucspect2}) adequately describes the fluctuation
behavior of bilayer membranes under both positive and negative
tensions.

\begin{figure}[!t]
  \centering
  \includegraphics[width=0.5\textwidth]{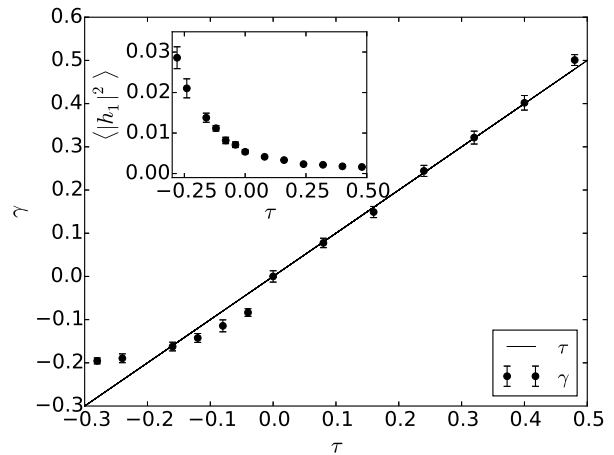}
  \caption{The fluctuation tension $\gamma$ as a function of the frame
  mechanical tension $\tau$. The solid line represents the equality
  $\gamma=\tau$ which is expected to hold for positive tensions. The
  inset shows the spectral intensity of the longest Fourier mode
  ($\vec n^2 = 1$), $\langle|h_1|^2\rangle$, as a function of $\tau$.}
  \label{fig:gamma}
\end{figure}

From the fitting curves, one can extract the values of $\kappa$ and
$\gamma\langle A_p\rangle$ as a function of $\tau$, and by
independently measuring the mean projected area, $\langle A_p\rangle$,
one can obtain the value of the fluctuation tension $\gamma$. Our
attempts to use $\kappa$ as a single fitting parameters by forcing
$\gamma=\tau$ resulted in poor fitting for negative tensions. The mean
projected area as a function of $\tau$ is plotted in the inset of
Fig.~\ref{fig:flucspect2}. The observed increase in $\langle
A_p\rangle$ with $\tau$ is anticipated and will be discussed in detail
later. Figure \ref{fig:gamma} depicts the fluctuation tension $\gamma$
as a function of $\tau$. The values reported in Fig.~\ref{fig:gamma}
are based on fitting analysis over the four longest fluctuation modes
(smallest wavenumbers), and the error bars represent the intervals
over which the fitting parameters, $\gamma$ and $\kappa$, can be
(mutually) varied while still producing reasonable fits up to the
accuracy of the computational results. For non-negative tensions, our
results agree very well with the relationship $\tau=\gamma$. As noted
above in section \ref{sec:negten}, there is no reason for this
equality to remain valid for negative tensions, and our analysis
summarized in Fig.~\ref{fig:gamma} reveals that, indeed,
$\gamma\neq\tau$ when the tensions are negative. Our results
demonstrate that $\gamma<\tau$ and, as also argued above, it is likely
that the more rapid decrease in $\gamma$ compared to $\tau$ is related
to the tendency of the membrane to form buckled configurations under
negative tensions.  The equality between $\gamma$ and $\tau$ is
regained for $\tau\simeq -0.15$, and then the trend changes, and
$\gamma$ becomes larger than $\tau$.

\begin{figure}[!t]
  \centering
  \includegraphics[width=0.5\textwidth]{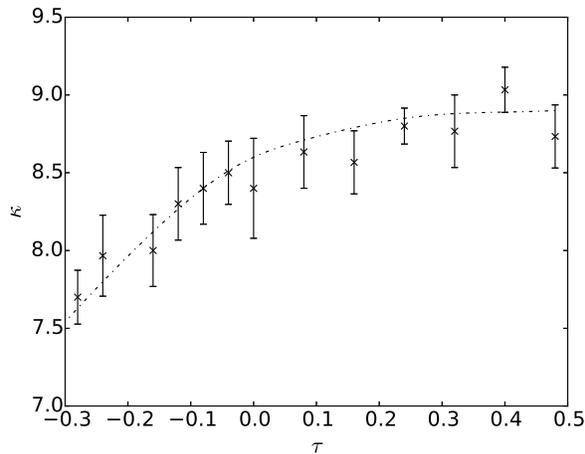}
  \caption{The bending rigidity $\kappa$ as a function of the frame
  tension $\tau$. The dotted-dashed line is a guide to the eye.}
  \label{fig:kappa}
\end{figure}

\begin{figure}[!t]
  \centering
  \includegraphics[width=0.5\textwidth]{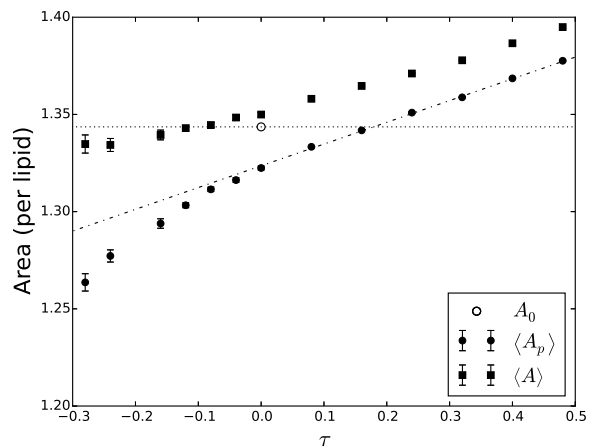}
  \caption{Measured area as a function of the frame tension
      $\tau$. Solids circles and squares denote, respectively, the
      results for the frame and total areas. The former was measured
      directly from the simulation, while the latter was derived from
      the computed data for the spectral intensity, by using
      Eq.~\ref{eq:totalarea}. The open circle marks the optimal area
      of a \emph{flat}\/ tensionless membrane, $A_0$. The dotted-dashed line 
      is a linear fit for the results for $\langle A_p\rangle$, while the 
      horizontal dotted line marks $A_0$. All areas plotted in the figure
      are normalized per lipid.}
  \label{fig:area}
\end{figure}

A closer inspection of the behavior of $\gamma$ depicted in
Fig.~\ref{fig:gamma} reveals that the $\gamma$ vs.~$\tau$ curve may be
divided into three regimes: (i) a linear $\gamma=\tau$ regime for
$\tau\geq 0$, (ii) a non-linear regime where $\gamma < \tau<0$ for
mildly negative frame tensions, and (iii) a plateau regime
($\gamma\sim {\rm const}$) for larger negative values of $\tau$.
Saturation of the negative tension for strongly compress membranes was
previously observed \cite{Otter:2005}, and we will return to the issue
later in this section when we discuss our results for the physical
area of the membranes (Fig.~\ref{fig:area}). The fluctuation tension
in Fig~\ref{fig:gamma} is extracted from Eq.~(\ref{eq:flucspect2}),
where it appears in the coefficient $4\pi^2\gamma\langle A_p\rangle$
of the $n^2$ term in the denominator. Naively, one may expect the
saturation of the fluctuation tension $\gamma$ to result in the
leveling-off of the fluctuation spectral intensity $|h_{\vec{n}}|^2$.
However, our computational results indicate that the amplitudes of the
normal undulations continue to grow for decreasing values of $\tau$,
as shown in the inset of Fig.~\ref{fig:gamma}. This apparent
discrepancy can be only partially resolved by the trend in $\langle
A_p\rangle$ whose value is reduced by about 10\% in the plateau regime
of $\gamma$. The main factor explaining the increase in the undulation
amplitude in the constant $\gamma$ regime is the decrease in the
effective bending modulus $\kappa$, the value of which is plotted in
Fig.~\ref{fig:kappa}. We remind here that $\kappa$ is not a material
but a thermodynamic quantity.  For a tensionless membrane, the thermal
undulations reduce (renormalize) the bending rigidity by $\Delta
k=-(3/4\pi) k_BT\ln(L/l)$, which is a small correction
\cite{kleinert}. For $\tau<0$, the amplitude of the fluctuations
increase and, therefore, this correction term should become larger (in
absolute value), which explains the drop in the value of $\kappa$ seen
in Fig.~\ref{fig:kappa}. In other words, just like the rapid decrease
in $\gamma$, reported above in Fig.~\ref{fig:gamma} for membranes
under negative tension, the reduction in $\kappa$ is also related to
the increasing thermal roughness of the membrane, and the tendency of
the membrane to form more buckled configurations.

The results of Figs.~\ref{fig:gamma} and \ref{fig:kappa} point to an
interesting difference between the elastic coefficients $\gamma$ and
$\kappa$.  The former decreases faster than $\tau$ for small negative
tensions and levels-off at large negative tensions. The latter
exhibits ``opposite'' behavior and remains fairly constant in the
small negative tension regime, and then decreases for strongly
compressed membranes. The crossover between the regimes occurs at
$\tau\simeq -0.15$. Some light may be shed on these observations by
the results of Fig.~\ref{fig:area} depicting the mean projected and
total areas as a function of $\tau$.  The results for the mean
projected area, $\langle A_p\rangle$, were measured directly from the
simulations, while the data for the mean total area, $\langle
A\rangle$, was calculated using Eq.~(\ref{eq:totalarea}). For
$\tau>0$, we observe a nearly linear dependence of both $\langle
A_p\rangle$ (see also the dotted-dashed line) and $\langle A\rangle$
on $\tau$. This behavior agrees very well with the experimental
results of Evans and Rawicz who also measured linear elastic response
of giant unilamellar vesicles (GUVs) under positive mechanical tension
\cite{evans}. We must point, however, to an important difference
between the origins of linear elasticity in GUVs and small bilayer
membranes. In both cases the linear elastic response is energetic in
nature and dominated by the area elasticity of the membrane, while the
entropy and bending energy of the height fluctuations play a secondary
role in the response to stretching. In GUVs, this happens after the
height fluctuations have been ironed by a very weak positive tension
scaling inversely with $A_p$. In small membranes, the height
fluctuations are not damped and, in fact, the simulation results in
Fig.~\ref{fig:area} reveal that the excess area ``stored'' in the
height fluctuations, $\langle A\rangle-\langle A_p\rangle$, decreases
only weakly with $\tau$. This implies that the entropy and bending
energy of small membranes do not vanish (as in GUVs under tension),
but simply exhibit relatively weak dependence on the frame tension
(and, therefore, contribute weakly to the elastic response).

\begin{figure}[!t]
  \centering \includegraphics[width=0.5\textwidth]{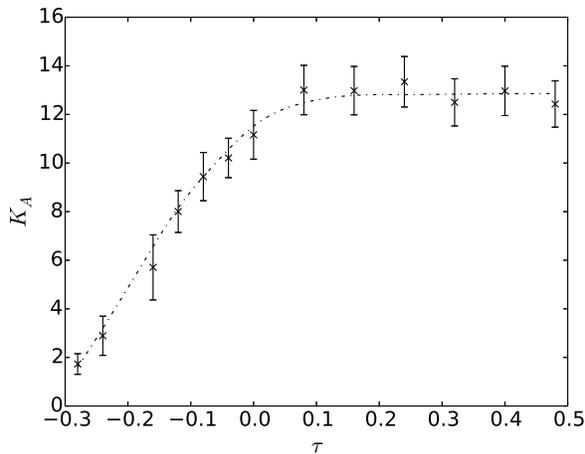}
  \caption{The stretch modulus $K_A$, measured from
  Eq.~(\ref{eq:compress1}), as a function of the frame tension
  $\tau$.The dotted-dashed line is a guide to the eye. }
  \label{fig:compress1}
\end{figure}

In addition to the simulations of fluctuating membranes, we also
simulated a tensionless ($\tau=0$) \emph{flat}\/ membranes by running
a MC code with moves allowing only local protrusions of lipids, but
completely suppressing the longer scale bending modes (i.e., ensuring
$h_{\vec{n}}=0$ for all $n$). For a flat membrane $A=A_p$. The
measured area of the flat tensionless membrane, $A_0$, is denoted by
the open circle and the horizontal dotted line in
Fig.~\ref{fig:area}. This is the area that minimizes the elastic
energy of the membrane.  Fig.~\ref{fig:area} provides an interesting
interpretation for the weak and strong negative tensions regimes. The
weak negative tension regime is essentially a continuation of the
positive tension regime. The mean area of a tensionless fluctuating
membrane is slightly larger than $A_0$, which implies that, in fact,
the membrane is stretched despite the negative mechanical tension.
Therefore, the area-dependent elastic energy continues to decrease
with $\tau$ into the weak negative tension regime. The strong negative
tension regime begins when the total physical area reaches
$A_0$. Since the membrane constitutes a dense fluid of lipids, it
cannot be much further compressed, and in order to maintain the total
area at $A_0$, more area must be expelled into the height
fluctuations. At this point, the elastic response becomes dominated by
the height fluctuations bending elasticity and entropy, and we begin
to observe a reduction in the effective bending modulus, $\kappa$,
instead of a reduction in the fluctuation tension $\gamma$.  The
saturation of the membrane physical area, and its correlation with
that of the surface tension, was previously reported
\cite{Otter:2005}. Here, we demonstrate that this occurs when$\left
\langle A \right \rangle$ reaches the value of $A_0$, which provided
an intuitive explanation to these observations.  Notice that the rapid
decrease in the projected area $\left \langle A_p \right \rangle$ with
$\tau$ in this regime is simultaneous to the increase in the projected
area fluctuations. The resulting rapid decrease in the effective
stretch modulus $K_A$, defined by Eq.~(\ref{eq:compress1}), is plotted
in Fig.~\ref{fig:compress1}. The vanishing of $K_A$ signals the onset
of mechanical instability \cite{comment3}.

\section{Discussion and Summary} 

\begin{figure}[!t]
  \centering
  \includegraphics[width=0.5\textwidth]{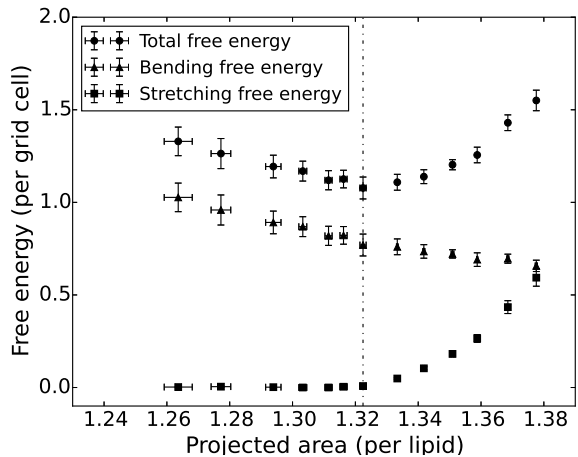}
  \caption{The bending, stretching, and total free energies (see
      definitions in the text) per grid cell, as a function of mean
    projected area $\left \langle A_p \right \rangle$ per lipid. The
    data for the total free energy has been shifted vertically by
    0.3
    for clarity. The vertical dotted-dashed line marks the measured
    projected area for $\tau=0$.}
  \label{fig:free_e}
\end{figure}

Based on our results, we identify two negative tension regimes with
the following features: (i) For weak negative tensions, the
fluctuation tension $\gamma$ drops somewhat faster than the mechanical
tension $\tau$, which is in contrast to the behavior observed for
positive tensions where $\gamma=\tau$. In this regime, the membrane is
still effectively stressed since the physical area $\langle
A\rangle>A_0$. (ii) For strong negative tensions, the fluctuation
tension saturates, but then the effective bending rigidity begins to
fall. Additionally, we also find that the total membrane area in this
regime reaches the optimal value of $A_0$, and does not continue to
drop much \cite{comment1}.

In the spirit of Eq.~(\ref{eq:helffreee}) for the elastic free energy
of positively stressed membranes, we can rationalize our observations
for negatively stressed membranes by writing the free energy as the
sum of two terms associated with stretching and bending. The former is
given by the quadratic form $F_{\rm stretch} = \left (1/2
\right)K_A\left [ \left \langle A \right \rangle - A_0 \right
]^2/A_0$, while the latter may be evaluated by $F_{\rm bend} =
\left(1/2 \right) \kappa \sum_{\vec n} n^4 \left \langle \left |
h_{\vec n} \right |^2\right \rangle$ \cite{comment2}. These two
contributions, and their sum are plotted in Fig.~\ref{fig:free_e} as a
function of $\langle A_p\rangle$.  The free energies in
Fig.~\ref{fig:free_e} provide insight into the computational results
in this work. First, we notice that the total free energy attains a
minimum at $\langle A_p \rangle/N\simeq 1.32$ (marked by the vertical
dotted-dashed line), which is the mean projected area measured for
$\tau=0$.  This must be the case since $\tau=\partial F/\partial
A_p$. Second, we notice that the bending free energy decreases with
$\tau$, which is expected since, upon stretching, the thermal bending
undulations tend to be suppressed. The stretching free energy increase
with $A_p$ in the weak negative and positive tensions regimes, i.e.,
when $\langle A\rangle>A_0$. Under strong negative tensions, $F_{\rm
stretch}$ vanishes, which is associated with the observation that the
total physical area remains at the optimal value and does not change
in this regime.

The crossover from bending- to stretching-dominated membrane
elasticity has also been observed in micropipette aspiration
experiments of GUVs. There are, however, several key differences
between the elastic behaviors of small and large membranes. Negative
mechanical tension destabilize the large bending modes of giant
membranes, while small membranes can withstand a (size-dependent)
negative tension that may be comparable in magnitude to the positive
rupture tension. In giant membranes, bending-dominated elasticity is
limited to extremely small positive tensions that are typically two
orders of magnitude smaller than the rupture tension. In small
membranes, the crossover from bending-dominated to
stretching-dominated elasticity is smoother and occurs at small
negative tensions. In other words, the stretching-dominated elasticity
regime extends into negative tensions, which stems from the fact that
at zero tension, the membrane is still slightly
stretched. Bending-dominated elasticity is observed at larger negative
tensions.  It is characterized by a decrease in the effective bending
rigidity and stretch modulus of the membrane that ultimately lead to
mechanical instability and membrane collapse.

This work was supported by the Israel Science Foundation, Grant
No.~1087.13.



\begin{thebibliography}{99}
\bibitem{comment4} Experimentally, the situation is more
complicated since one deals with closed vesicles, whose projected area
is not as well defined as the projected area of open bilayers with
periodic boundary conditions. The thermodynamics of fluid vesicles was
carefully analyzed in ref.~{\protect \cite{diamant}}
\bibitem{comment3} It it worthwhile to reemphasize the coupling
between the statistics of the projected area and the height
fluctuations, resulting from the saturation of the total area. Thus,
the decrease in $K_A$ does not reflect a change in the material
stiffness of the bilayer membrane, but is simply just another
reflection of the fact that it becomes increasingly buckled in
response to $\tau$ becoming more negative.
\bibitem{comment1} The determination of the physical area is based on
Eq.~(\ref{eq:totalarea}), which is a Taylor expansion of $\langle
A\rangle$ for weakly fluctuating membranes. The slight decrease in
$\langle A\rangle$ observed for large negative tensions (i.e., when
the membrane undulates more strongly) may be an artifact of this
approximation.
\bibitem{comment2} The elastic free energy is obviously a function of the
projected area $F=F(A_p)$. In the expressions here, the dependence on
$A_p$ is implicit through the dependence of the statistical averages
of $A$ and $|h_n|^2$, and through the values of the renormalized
elastic moduli.
\end{thebibliography}
\end{document}